\documentclass[prl,twocolumn,preprintnumbers,amsmath,amssymb]{revtex4}
\usepackage{graphicx}

\begin{document}
\title{  Super-radiance, Berry phase, Photon phase diffusion and Number squeezed state in the $ U(1) $ Dicke ( Tavis-Cummings ) model   }
%\title{ Super-radiance, Berry phase, Photon phase diffusion and Number squeezed state of cold atoms or superconducting qubits inside a cavity   }
\author{ Jinwu Ye$^{1,2}$ and  CunLin Zhang $^{1}$   }
\affiliation{$^{1}$ Beijing Key Laboratory for Terahertz
Spectroscopy and Imaging, Key Laboratory of Terahertz
Optoelectronics, Ministry of Education, Department of Physics,
Capital Normal University, Beijing, 100048 China  \\
$^{2}$ Department
of Physics and Astronomy, Mississippi State University,
%P. O. Box 5167, Mississippi State,
MS, 39762, USA  }
\date{\today }

\begin{abstract}
  Recently, strong coupling regimes of  superconducting qubits or quantum dots inside a micro-wave circuit cavity and BEC atoms inside an optical cavity
  were achieved  experimentally. The strong coupling regimes in these systems were described by the Dicke model.
  Here, we solve the Dicke model by a $ 1/N $ expansion.
  In the normal state, we find a $ \sqrt{N} $ behavior of the collective Rabi splitting. In the superradiant phase, we identify an important Berry phase
  term which has dramatic effects on both the ground state and
  the excitation spectra of the strongly interacting system. The single photon excitation spectrum has a low energy quantum
  phase diffusion  mode in {\sl imaginary time } with a large spectral weight
  and also a high energy optical mode  with a low spectral weight. The photons are in a number squeezed state which
  may have wide applications in high sensitive measurements and quantum information
  processing. Comparisons with exact diagonization studies are made.
  Possible experimental schemes to realize the superradiant phase are briefly discussed.
%  By analyzing the Berry phase effect on this
%  quantum phase diffusion mode, we work out many remarkable experimental consequences of this  mode such as its low frequency, high spectral weight,
%  periodicity in one photon number increase,  photon number squeezing properties and photon statistics.
%  The photons in this quantum phase diffusion mode  are in a number squeezed state  with much enhanced signal/noise ratio which
%  may have wide applications in the field of high resolution, high sensitive measurement and also in quantum information processing.
\end{abstract}

\maketitle

% The $ U(1) $  Dicke ( Tavis-Cummings ) model describes finite number of cavity photon
% modes coupled to an assembly of $ N $ atoms with the same strength.

%% Since the Dicke models were proposed in 1968, they have been studied by various methods,
%% most of them focusing on the thermodynamic limit $ N=\infty $ inside a perfect cavity with $ Q= \infty  $.

% In the thermodynamic limit $ N =\infty $, it was known that as the collective atom-photon coupling strength
% increases, the system will evolve from the normal phase into a super-radiant phase where the $ N $ atoms radiate collectively.
% Recently, several experiments successfully achieved the strong coupling of a BEC of $ N \sim 10^{5}
% $ $ ^{87} Rb $ atoms to the photons inside an ultrahigh-finess optical cavity.
% In the strong coupling regime, the individual coupling strength  between
% the ( artificial ) atoms and field is larger than the spontaneous decay rate of the
% upper level and also the dissipation rate of the intra-cavity field.
% The strong coupling regime was also achieved  with  superconducting qubits, quantum dots or electron spins inside a micro-wave circuit cavity.
% These experiments are precisely described by the $ U(1) $ Dicke model.

%% realized the normal phase of the $ U(1) $ Dicke model.
%The importance of various kinds of Dicke
%models in quantum optics ranks the same  as the boson Hubbard
%models, Fermionic Hubbard models,  Heisenberg models in strongly
%correlated systems and the Ising model  in Statistical  mechanics.

   Recently, several experiments \cite{qedbec}
   successfully achieved the strong coupling of a BEC of $ N \sim 10^{5} $ $ ^{87}Rb $ atoms to
   the photons inside an ultrahigh-finesse optical cavity.
 In parallel, strong coupling regime was also achieved  with artificial atoms such as
 superconducting qubits inside micro-wave circuit cavity and  quantum dots inside a semi-conductor micro-cavity system \cite{circuit}.
  In these experiments,  the individual maximum coupling strength $ \tilde{g} $ between
 the ( artificial ) atoms and field is larger than the spontaneous decay rate of the
 upper state $ \gamma $ and the intra-cavity field decay rate $ \kappa $.
% The relevant Fibre-based Fabry-Perot (FFP) cavity parameters
%   are $ ( \tilde{g}, \kappa, \gamma )= 2 \pi \times ( 10.6, 1.3, 3.0) MHz $ with  very
%   high collective cooperativity $ C = N \tilde{g}^{2}/ \kappa \gamma \sim 10^{6} $.
 The collective Rabi splitting was found to scale as $ \sqrt{N} $.
 All these  systems are described by the Dicke model\cite{dicke} Eqn.\ref{dickeu1} where a
 single mode of photons coupled to an assembly of $ N $ atoms with the same coupling strength $ \tilde{g} $.

The importance of various kinds of Dicke models in quantum optics
ranks the same as the boson Hubbard model, Fermionic Hubbard model,
Heisenberg model in strongly correlated  systems and the Ising model
in Statistical mechanics.
 Since  the Dicke model was proposed in 1954, it was solved in the thermodynamic limit $ N= \infty  $  by various methods \cite{dicke1,popov,staircase,zero,chaos}.
 It was found that when the collective atom-photon coupling strength is sufficiently large ( Fig.1 ),
 the system gets into a new phase called super-radiant phase where there are large number of inverted atoms and also large number photons
 in the system's ground state \cite{noneq}.
% the $ N $ atoms  interact
% with photons  collectively and radiate collectively \cite{noneq}.
%   its properties was still far from being understood.
%   In this proposal, we will study fantastic and important physics of various kinds of Dicke models, especially
%   in the super-radiant coupling regime and their experimental implications.
  However, so far, there are only a few very preliminary exact diagonization (ED) study on Dicke models at finite  $ N $ \cite{staircase,chaos},
  its underlying physics remains unexplored \cite{bethe}.
  It is known that any real symmetry breaking  happens only at the
  thermodynamic limit $ N \rightarrow \infty $, so in principle, there is no real symmetry breaking, so no real super-radiant phase at any finite $ N $.
  But there is a very important new physics for a finite system $ N $  called
  quantum phase diffusion in  { \sl imaginary time }  at finite $ N $  for a continuous symmetry breaking
  ground state at $ N =\infty $. The quantum  phase diffusion process
  in a finite system is as fundamental and  universal as symmetry breaking in an infinite system.
  Here, we will explore the quantum phase diffusion process of the Dicke model by a $ 1/N $
  expansion. We determine the ground state and single photon excitation
  spectrum in both normal and superradiant phase.
  In the normal state, we find a $ \sqrt{N} $ behavior of the collective Rabi splitting in the single photon excitation spectrum consistent with
  the experimental data  and also determine the corresponding spectral weights.
  In the superradiant phase, we identify a Berry phase  term which has dramatic effects on both ground state and
  the excitation spectra.
  The single photon excitation spectrum has a very low energy quantum  phase diffusion
  mode $ E_D $ with a high spectral weight and also a high energy optical mode $ E_{o} $ with a low spectral weight.
  Their energies and the corresponding spectral weights are calculated.
  The photons are in a number squeezed state. The squeezing parameter ( namely, the Mandel $ Q_M $ factor ) is determined.
  It is the Berry phase which leads to the "Sidney Opera " shape  in the single photon excitation spectrum
  and the consecutive plateaus in photon numbers in Fig.1. The Berry
  phase is also vital to make quantitative comparisons between the
  analytical results in this paper and the very preliminary ED results
  in \cite{staircase}  and much more extensive ED in \cite{long}.
   Being very strong in intensity and has much enhanced signal/noise ratio, the number squeezed state from the superradiant phase
   may have wide applications in quantum information processing \cite{subrev} and also
   in the field of high resolution and high sensitive measurement \cite{ligo}.
%  So far, this may be the only model where one can calculate the quantum phase diffusion constant $ D $ exactly.
%  The cavity systems maybe the most effective ones to detect the very important quantum effects of the Berry phase and phase diffusion process
%  in a finite system which is as fundamental and  universal as symmetry breaking in an infinite system.
   Several experimental schemes to realize the superradiant phase  of the $ U(1) $ Dicke model briefly discussed.
\begin{figure}
\includegraphics[width=7cm]{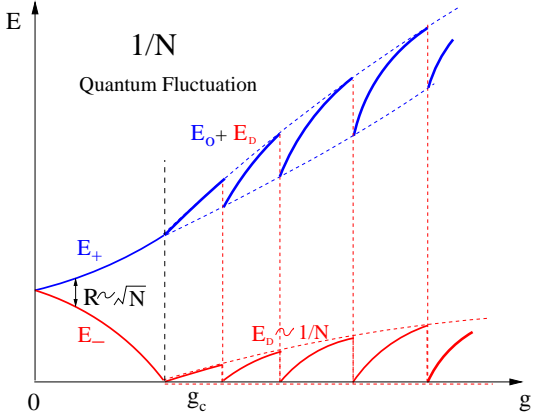}
\caption{ ( Color online ) The single particle excitation spectrum
 E ( in energy unit ) of the $ U(1) $ Dicke ( Tavis-Cummings ) model Eqn.\ref{dickeu1} versus the collective atom-photon
 coupling strength $ g $ ( in energy unit ).
 The critical coupling  is $ g_c= \sqrt{ \omega_c \omega_a} $.
 In the normal state $ g < g_{c} $,  the single photon spectrum $  S( \omega )
 $ has two peaks at the lower branch $ E_{-} $
 and the upper branch $ E_{+} $ with spectral weights $ c_{-} \sim 1 $ and $ c_{+} \sim 1 $ respectively.
 The collective Rabi splitting $ R= E_{+}- E_{-}  \sim \tilde{g} \sqrt{N} $  was measured in \cite{qedbec}.
 In the super-radiant phase  $ g > g_{c} $, the zero mode indicated by the dashed red line at $ g > g_c $
 is lifted to the phase diffusion mode $ E_{D} $ in Eqn.\ref{ed}  by the $ 1/N $ correction.
 The single photon spectrum  $  S( \omega ) $
 has a peak at the low frequency phase  diffusion  mode $ E_D \sim \omega_c/N ( \frac{1}{2} + \alpha ) $ with the spectral weight $  c_{D}\sim N \lambda^{2} \sim N $ and
  a high frequency optical mode $ E^{t}_{o}= E_o+ E_D  $ with
  a spectral weight $ c_{o} \sim 1 $. It is the Berry phase (  $
-1/2<\alpha < 1/2 $  ) effects  which lead to the " Sidney Opera "
shape of the single photon energy spectrum. } \label{fig1}
\end{figure}

  In the $ U(1) $ Dicke model \cite{dicke}, a single mode of photons couple to $
  N $ two level atoms with the same coupling  constant $ \tilde{g} $. The two level
  atoms can be expressed in terms of 3 Pauli matrices $ \sigma_{\alpha},
  \alpha=1,2,3 $. Under the Rotating Wave (RW) approximation, the $ U(1) $ Dicke model can be written as:
\begin{equation}
  H_{U(1)}  =  \omega_{c} a^{\dagger} a +  \frac{\omega_{a}}{2} \sum^{N}_{i=1} \sigma^{z}_{i}
   + \frac{g}{\sqrt{N}}\sum^{N}_{i=1} ( a^{\dagger} \sigma^{-}_{i} + h.c. )
\label{dickeu1}
\end{equation}
   where the $ \omega_c, \omega_a $ are the cavity photon frequency, the energy difference of the two atomic levels  respectively,
   the  $  g= \sqrt{N} \tilde{g} $ is the collective photon-atom coupling (  $  \tilde{g} $ is the individual photon-atom coupling ),
   the cavity mode $ a $ could be any one of the two orthogonal polarizations of $
   TEM_{00} $ cavity modes in \cite{qedbec}.
%   One can represent the spin operators in terms of
%   two fermions $ c_{\alpha}, \alpha=0,1 $  subject to the local constraint  $ \sum_{ \lambda } c^{\dagger}_{i \lambda} c_{i
%   \lambda}=1 $: $ \sigma^{z}_{i}=  c^{\dagger}_{i 1} c_{i 1}-c^{\dagger}_{i 0} c_{i
%   0}, \sigma^{+}_{i}= c^{\dagger}_{i0} c_{i1}, \sigma^{-}_{i}= c^{\dagger}_{i1} c_{i0} $.
   One can also add the atom-atom interaction $ H_{at-at} $ to the
   Eqn.\ref{dickeu1}.  Because $ H_{at-at} $ does not change the symmetry of the model,
   we expect all the results achieved in the paper remain qualitatively valid. The Hamiltonian Eqn.\ref{dickeu1} has the $ U(1) $ symmetry $ a
   \rightarrow  a  e^{ i \theta}, \sigma^{-} \rightarrow \sigma^{-} e^{ i \theta} $. In the normal phase, $ \langle a \rangle =0 $, the $
   U(1) $ symmetry is respected. In the super-radiant phase,
   $ \langle a \rangle \neq 0 $, the $ U(1) $ symmetry is spontaneously broken.
   The model Eqn.\ref{dickeu1}  was studied by a large $ N $ expansion in Ref.\cite{popov,zero}.
   However, they did not extract any important physics at the order of $1/N $.
   In this paper, we will show that by carefully analyzing the effects of the zero mode in the super-radiant phase, one can extract
   the most important physics of quantum phase diffusion at the order $ 1/N $.
   In the large $ N $ expansion in the magnetic systems \cite{spn}, $ N $ is the order of
   the magnetic symmetry group  $ O(N), SU(N), Sp(2N) $ with $ N=3,2,1 $ respectively.
   However here $ N \sim 10^{5} $ is the number of atoms, so $ 1/N \sim 10^{-5} $
   expansion is quite accurate.

   Following standard large $ N $ techniques developed in \cite{spn}, after re-scaling the
   photon field $ a \rightarrow \sqrt{N} a $,  integrating out
%   the fermions subject to the local constraint,
   the spin degree of freedoms, one can get an effective action  $  S_{eff} [a] $  in terms of the
   photon field only, then  perform a large $ N $ expansion. At $ N \rightarrow \infty $,
   the photon mean field value $ \langle a( \tau ) \rangle = \lambda $ is determined \cite{dicke,dicke1,staircase,popov}
   by the saddle point equation:
%    $ \frac{ \delta S_{eff} [a] }{ \delta a^{*} }=0 $
%   given in \cite{dicke,dicke1,dicke2,staircase,popov,popovbook}.
  $ \omega_{c}  \lambda =  g^{2} \lambda \frac{ \tanh \beta E }{ 2 E } $
   where $ E= \sqrt{ (\frac{\omega_a}{2})^{2} + g^{2} \lambda^{2} } $ and $ \beta=1/k_{B} T $ is the inverse temperature.
   At $ T=0 $, in the normal phase $ g < g_{c}= \sqrt { \omega_a \omega_c }$, $ \lambda=0 $, the photon number $ n/N =0 $;
   in the super-radiant phase $ g > g_{c} $, $ \lambda \neq 0 $,  the photon number $ n/N = \lambda^{2} \sim g- g_c   $ when $ g $ is slightly above $ g_c $.
   Its phase diagram and photon number at $ N=\infty $ and finite $ N $ is shown in Fig.1 and 2 respectively.

    At finite $ N $, writing  $ a= \lambda + \psi $ where $ \psi $ describes the photon fluctuation around its mean field value
   $ \lambda $, one can expand the $ S[a] = S_0 + S_2+ S_3 +\cdots $  to second order  \cite{zero} in $ \psi $:
\begin{eqnarray}
      S_{2} [ \bar{\psi}, \psi ] & = &  \frac{N}{ 2 \beta} \sum_{i
      \omega  } (  \bar{\psi} ( \omega ), \psi( - \omega ) )  G^{-1}  \left ( \begin{array}{c}
   \psi( \omega )   \\
   \bar{\psi} ( - \omega )    \\
   \end{array}   \right )     \nonumber  \\
   G^{-1} & = & \left ( \begin{array}{cc}
     K_{1}   &  K_{2}  \\
     K^{*}_{2}  &  K^{*}_{1}   \\
   \end{array}   \right )
\label{second}
\end{eqnarray}
   where the explicit expressions of the $ K_1, K_2 $ are given in \cite{zero}, but are not needed in the following.

      In the normal phase $ g < g_{c} $, $ \lambda=0 $.
      One can see the normal Green function  at $ T=0 $ in Eqn.\ref{second} has two poles
      $ E_{\pm} = \frac{ ( \omega_c + \omega_a) \pm \sqrt{ ( \omega_c-\omega_a)^{2}+ 4 g^{2} } }{2} $
      with the spectral weights $ c_{+}= \frac{ E_{+} -\omega_a }{ E_{+} - E_{-} }, c_{-}= \frac{ \omega_a -E_{-} }{ E_{+} - E_{-} } $  ( Fig.1 ).
      After the analytic continuation $ \tau \rightarrow it $, the one photon Green function at $ T=0 $ takes:
\begin{equation}
 \langle a(t) a^{\dagger}( 0 ) \rangle_{N}  \sim  c_{-} e^{-i E_{-} t} + c_{+} e^{- i E_{+} t}
\label{onephotonnormal}
\end{equation}
      where we also put back the re-scaling factor of the photon field $ a \rightarrow \sqrt{N} a $. It
      leads to the two peaks at the two poles $  E_{\pm} $ in the single photon energy spectrum  shown in the Fig.1.
      At the resonance $ \Delta_c= \omega_c-\omega_a =0 $, the collective Rabi splitting
      $ E_{+} - E_{-}= 4g = 4 \sqrt{N} \tilde{g} \sim \sqrt{N}  $ shown in Fig.1 was measured in \cite{qedbec}.
      Note that due to $ \tilde{g}_{\sigma_{+}} >  \tilde{g}_{\sigma_{-}} $ in\cite{qedbec}, the coefficients of the $ \sqrt{N} $ are different for the
      two different polarizations.
%      Here, we re-derived this important result from the $ 1/N $ expansion of the Dicke model which
%      may bring additional physical insights into the intuitive arguments in\cite{qedbec}.
      The intensity ratio of the two peaks  $ \frac{c_{+}}{c_{-}} = \frac{ E_{+} -\omega_{a} }{ \omega_a -E_{-}  } $
      seems has not been measured yet  in \cite{qedbec}.

      However, at the super-radiant phase $ g > g_{c} $, $ \lambda \neq 0 $.
      the anomalous term $  K_2 \neq 0 $ and $ |K_{1} |^{2} - |K_{2} |^{2} $  contains a zero mode shown as a red dashed
      line in Fig.1,  in addition to the pole at a high frequency
      $ E_{o} = \sqrt{ ( \omega_c + \omega_a)^{2} + 4 g^2 \lambda^{2} } $ shown as the blue dashed line in Fig.1. This "zero " mode is nothing but the " Goldstone "
      mode due to the global $ U(1) $ symmetry breaking in the super-radiant phase.
      The important physics behind this "zero"
      mode was never addressed in the previous literatures  \cite{popov,zero}. Here we will explore the remarkable properties of this "zero" mode.
      Because of the infra-red divergences from this zero mode, the $ 1/N $ expansion in
      the Cartesian coordinates need to be summed to infinite orders to lead to a finite physical result.
      It turns out that the non-perturbative effects of the zero mode  can
      be more easily analyzed in the polar coordinate ( or phase representation )  by writing $ a= \lambda + \psi_{1} + i\psi_{2} = \sqrt{ \lambda^{2} +
      \delta \rho } e^{ i \theta} $, then to linear order in $ \delta \rho $ and $ \theta $:  $ \psi_{2} = \lambda \theta  $.
      In Eqn.\ref{second}, by integrating out the massive $ \psi_{1} $ mode, using  $ \psi_{2} = \lambda \theta  $,
      also paying a special attention to the Berry phase term  \cite{bertsubir}  coming from the angle variable $ \theta $,
      one can show that the dynamics of the  phase $ \theta  $ is
      given by:
\begin{equation}
   S_{2} [ \theta ] =  i N \lambda^{2} \partial_{\tau} \theta + \frac{N}{ 2 \beta} \sum_{i
      \omega  } \frac{ 2 \lambda^{2} \omega^{2}  ( \omega^{2} + E^{2}_{o} ) }{  \omega_{c} (\omega^{2} + 4
      g^{2} \lambda^{2} ) } |\theta( \omega ) |^{2}
\label{angle}
\end{equation}
      In the following, we will discuss the the zero mode and the optical mode respectively.

      In the low frequency $ \omega \ll E_o  $ limit where the magnitude fluctuations can be dropped,
      Eqn.\ref{angle} reduces to:
\begin{equation}
     {\cal L}_{PD} [ \theta ]  =  i N \lambda^{2} \partial_{\tau} \theta +  \frac{1}{2 D }  ( \partial_{\tau} \theta )^{2}
                   =  \frac{1}{2 D }  ( \partial_{\tau} \theta + i \alpha D )^{2}
\label{diffu1}
\end{equation}
     with the quantum phase diffusion constant $  D = \frac{ 2 \omega_c g^{2} }{ E^{2}_{o} N }
     $. In the Eqn.\ref{diffu1}, we have denoted
     $  N \lambda^{2} = N_{0} + \alpha $ where $ N_0=[ N \lambda^{2}]  $ is the {\sl closest } integer to $ N \lambda^{2} $, so $ -1/2 < \alpha < 1/2
     $.

      The corresponding quantum phase diffusion Hamiltonian is:
\begin{equation}
      H_{PD} [ \theta ]  =  \frac{D}{2} ( \delta N_{ph} - \alpha  )^{2}
\label{diffu1h}
\end{equation}
     where $ \delta{N}_{ph}= N_{ph}- N_0 $ is the photon number fluctuation
     around its ground state value $  N_0 $ and is conjugate to the phase $ \theta $:
     $ [ \theta, \delta N_{ph} ]= i \hbar $. In fact,
     Eqn.\ref{diffu1h} can be considered as the Hamiltonian of a particle moving along a
     ring with a very large inertial of moment $ I=1/D $ subject to
     a fractional flux $ f= \phi/\phi_{0} = \alpha $.

     In Eqn.\ref{diffu1},  after defining $ \tilde{\theta}(\tau)=  \theta(\tau) + i \alpha D \tau $, one can easily show that
\begin{equation}
     \langle ( \tilde{\theta}(\tau) -\tilde{\theta} (0))^{2} \rangle
    = 2 D  \int \frac{ d \omega }{ 2 \pi }  \frac{1- e^{i \omega \tau}
    }{ \omega^{2} }= D |\tau|
\label{dtau}
\end{equation}
    which is a phase diffusion in {\sl imaginary time $ \tau $ } \cite{collapse,laser} with the phase diffusion constant $ D $.
    Only in the thermodynamic limit $ N \rightarrow \infty $, a
    state with a given initial phase will stick to this phase as the time evolves, so we have a
    spontaneously broken $ U(1) $ symmetry. However, for any finite
    $ N $, the initial phase  has to diffuse with
    the phase diffusion constant $ D \sim 1/N $.
    The diffusion time scale in the {\sl imaginary time } beyond which there is no more phase
    coherence is $ \tau_{D}=1/D \sim  N/\omega_c $ which is finite
    for any finite $ N $. This can also be called phase
    "de-coherence" time in the {\sl imaginary time } \cite{laser}.

     From Eqn.\ref{diffu1}, it is easy to see that the gapless nature of the phase diffusion mode in Eqn.\ref{diffu1} leads
     to the vanishing of the order parameter inside the super-radiant phase:
\begin{equation}
      \langle a \rangle = 0
\label{average}
\end{equation}
     So the $ U(1) $ symmetry is restored by the phase diffusion.
%     In the language of the quantum optics, the phase noise
%     is well above the standard quantum limit.
     From Eqn.\ref{diffu1}, after doing the analytic  continuation $ \tau \rightarrow i t
     $, we can get:
\begin{equation}
 \langle a^{\dagger}( t) a( 0 ) \rangle_{S} =  N \lambda^{2} e^{-i (\frac{1}{2} +\alpha) D t}
\label{onephotondicke}
\end{equation}
     where we also put back the re-scaling factor of the photon field $ a \rightarrow \sqrt{N} a $.
%     Putting $ t=0 $ in Eqn.\ref{onephotonnormal} and
%     \ref{onephotondicke} lead to the average photon number at $ N <
%     \infty $ in the ground state of the normal $ n/N=0 $ and the super-radiant phase $ n/N = \lambda^{2} $
%     respectively shown in Fig.2.
     It is also easy to see that $ \langle a( t) a( 0 ) \rangle_{S} =0 $, so there is no quadrature squeezing anymore at any finite $ N $ \cite{squeezing}.
     In fact, all these results can also be achieved by using the
     Hamiltonian Eqn.\ref{diffu1h}.

     Eqn.\ref{onephotondicke} leads to the result that
     the energy of the "zero energy mode " ( Goldstone mode ) at $ N=\infty $ was "lifted"
     to a  quantum phase "diffusion " mode at any finite $ N $  with a finite  small positive frequency \cite{laser}:
\begin{equation}
      E_{D} = ( \frac{1}{2} + \alpha ) D =  \frac{ 2 \omega_c g^{2} ( \frac{1}{2} + \alpha ) }{ [ ( \omega_c + \omega_a)^{2} + 4 g^2 \lambda^{2}] N }   \sim \omega_c/N
\label{ed}
\end{equation}
     It is the Berry phase effect which
     leads to the periodic jumps in the Fig.1.
     The Fourier transform of Eqn.\ref{onephotondicke} leads to the
     Fluorescence spectrum  $ S( \omega )  =   N \lambda^{2} \delta( \omega- E_{D} ) $  with the spectral weight
     $ c^{0}_{D} = N \lambda^{2} \sim N $.

\begin{figure}
\includegraphics[width=6cm]{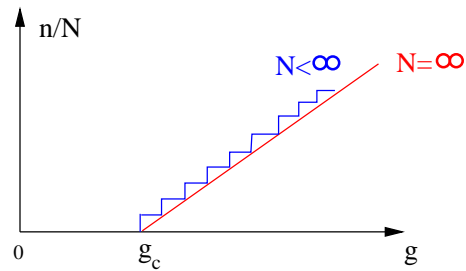}
\caption{ ( Color online ) The average photon number in the ground
state at $ N = \infty $ and at a finite $ N < \infty $ versus the
collective atom-photon coupling strength $ g $ ( in energy unit ).
The photon is very close to be the Fock state as shown by the steps
in the superradiant phase at $ N < \infty $. It is the Berry phase (
$ -1/2<\alpha < 1/2 $  ) effects  which lead to the  steps in the
photon number. } \label{fig2}
\end{figure}

     Now we study the photon statistics. If neglecting the magnitude fluctuation,  the quantum phase diffusion Hamiltonian Eqn.\ref{diffu1h}
     shows that the ground state is a photon Fock state with eigenvalue $ N_0 $
     which jumps by 1 in all the plateaus ending at $ \alpha=1/2 $ in the Fig.2.
%     This is consistent with the fact that the phase noise is very large due to the phase diffusion process in Eqn.\ref{dtau}.
     Now we incorporate the magnitude fluctuation. In the Eqn.\ref{second}, by integrating out the imaginary part $ \psi_{2}( \omega ) $ and  using $
    \psi_1= \delta \rho/2 \lambda $, one can get the effective action for the magnitude
    fluctuations
\begin{equation}
      {\cal L }_{2}(\delta \rho  )=  N \frac{  \omega^{2}+ E^{2}_{o} }{ 8 \lambda^{2} \omega_c } | \delta \rho ( \omega ) |^{2}
\label{mag}
\end{equation}
     where we find the Mandel factor $   Q_{M}= \frac{ \langle ( \delta N )^{2}
  \rangle - \langle N \rangle }{  \langle N \rangle }=-1+ \frac{ 2 \omega_c }  {E_o} $
  so the deviation from the Fock state at any given plateau in Fig.2 is given by  $ \frac{ 2 \omega_c } {E_o} $.
  Because $  \frac{ 2 \omega_c }{ E_o} \ll 1 $ in the $ g\gg g_c $ limit, it is very close to be a Fock state.
  This is a highly non-classical state with Sub-Poissonian  photon statistics.
  It has very strong signal $    \langle N_{ph} \rangle = N \lambda^{2} $, but nearly no photon number
  noise, so it has  a very large signal to noise ratio which could be crucial for quantum information processing \cite{subrev}
  and also in the field of high resolution and high sensitive measurement \cite{ligo}.

   Using the polar  representation $ a =\sqrt{ \lambda^{2} + \delta \rho } e^{ i \theta }  \sim \lambda e^{ i \theta } +
   \frac{ \delta \rho }{ 2 \lambda } e^{ i \theta }  + O(1/N) $, one can  evaluate the photon correlation function:
\begin{eqnarray}
  \langle {\cal T } a^{\dagger}( \tau ) a(0) \rangle  =   N \lambda^{2}
  \langle e^{-i ( \theta( \tau) - \theta (0) ) } \rangle~~~~~~~~~~~~   \nonumber
  \\
 ~~~~~~~~ + \frac{ N
  }{ 4 \lambda^{2} } \langle \delta \rho( \tau) \delta \rho(0) \rangle
  \langle e^{-i ( \theta( \tau) - \theta (0) ) } \rangle + O(1/N)
\label{wick}
\end{eqnarray}
     where the $ {\cal T} $ means imaginary time ordered. By evaluating the first
     and second term from Eqns.\ref{angle} and \ref{mag}, we can identify not only the quantum phase diffusion
     mode $ E_D =D ( \frac{1}{2} + \alpha ) $ with the corresponding spectral weight $  c_{D}=N \lambda^{2}- \frac{ \omega_c ( \omega_c + \omega_a)^{2} }{4  E^{3}_{o} } $,
     but also the optical mode $ E^{t}_o=  E_{o}+E_{D} $ with the corresponding  spectral weight
     $ c_o= \frac{ \omega_c ( \omega_c + \omega_a)^{2} }{4  E^{3}_{o} } +\frac{ 2 \omega_c }{ E_o} $ ( Fig.1 ).
     Note that  $ E^{t}_o- E_{D}=  E_{o} $ in Fig.1 is independent of the
     Berry phase $ \alpha $.  So the total energy in the optical frequency peak $ \sim E^{t}_{o} \times c_{o} \sim \omega_c $ is comparable to that
     in the phase diffusion mode $ \sim E_{D} \times c_{D} \sim \omega_c/N \times N \sim \omega_c $.

  If one introduce the total "spin"  of the $ N $ two level
  atoms $ J^{z}= \sum_{i} \sigma^{z}_{i}, J^{+}= \sum_{i}  \sigma^{+}_{i}, J^{-}= \sum_{i} \sigma^{-}_{i} $
  and confine the Hilbert space only to $ J=N/2 $, then Eqn.\ref{dickeu1} can be simplified to the $ J-U(1) $ Dicke model
  which was studied  by an exact diagonization (ED) in \cite{staircase}.
  The authors in \cite{staircase} found that there are a series of ground state energy level crossings as the $ g $ gets
  into the super-radiant regime and interpreted them as consecutive "quantum phase transitions ". But they did not
  study any excited states.  In \cite{long}, we performed a much more extensive ED study directly on the $
  U(1) $ Dicke model Eqn.\ref{dickeu1} and not only calculated the ground states, but also all the excited energy levels.
  We also identified a series of ground state energy level crossings in the super-radiant regime.
  By comparing with the analytic results achieved in this paper, we found all these ground states crossings
  are precisely due to the periodic changes of the Berry phase $ \alpha $ in Eqns.\ref{angle},\ref{diffu1},\ref{diffu1h}.
  They are not consecutive "quantum phase-like transitions " as
  claimed in \cite{staircase}.
%  As explained in the introduction, there are no  phase transitions anymore in any finite number of atoms.
  We also found one to one quantitative matches between the low energy phase
  diffusion mode $ E_D $, also the high energy optical mode $ E^{t}_o=  E_{o}+E_{D} $ and the
  excited levels found by the ED in \cite{long} at $ N $ as small $ N=5 $.
  The complete comparisons will be presented in \cite{long}.

  It remains experimentally challenging to move into the superradiant  regime which requires the collective photon-atom coupling
  $  g= \sqrt{N} \tilde{g} >  g_{c}= \sqrt{ \omega_{c} \omega_{a} } \sim 2 \pi \times 10^{5} GHz $ for the optical cavity used in \cite{qedbec}.
  The collective Rabi splitting  $ \sqrt{N} \tilde{g} \sim 20 GHz $ in \cite{qedbec}
  is still much smaller than $ \omega_{c} =  2 \pi \times 10^{5} GHz $,
  so not even close to the superradiant regime in Fig.1.
  It was proposed in \cite{dickej,orbitalthermal,orbital} that the super-radiant regime can be realized
  by using a cavity-plus-laser-mediated  Raman transitions between a pair of stable atomic ground states, therefore also suppress the spontaneous emission  $ \gamma $.
  All the parameters in Eqn.\ref{dickeu1}
  can be controlled by the external  laser frequencies and intensities, so the
  characteristics energy scales in the effective two level atoms  are no longer those of optical photons and dipole coupling,
  but those  associated with Raman transition rates and light shifts.
  Indeed, using this scheme, the super-radiant phase in the $ Z_2 $ Dicke model \cite{chaos,dickej} was reached
  by using both thermal atoms \cite{orbitalthermal} and the cold atoms in the BEC
  \cite{orbital}. We expect this scheme may also be used to realize
  the super-radiant phase of the $ U(1) $ Dicke model shown in Fig.1.
  Because the microwave circuit cavity has much lower cavity frequency and the individual photon-qubit $ \tilde{g} $
  can also be made very large, so  the superradiant phase could also be realized in
  superconducting qubits or quantum dots inside a circuit cavity in the future.
%  where the effective two "atomic" levels are the two momentum states of the cold atoms
%  in the optical lattice formed by the cavity field and the off-resonant  Laser pumping {\sl transversely } to the cavity axis.
  In the experiments, there is also a weak dissipation $ \kappa \ll g  $.  In a future publication,
  following the procedures in \cite{squeezing}, we will study the effects of $ \kappa $ on the number squeezed state.

%  By following the procedures in \cite{squeezing}, we
%  expect it will lead to a very small shift and broadening
%  to the florescence spectrum in Fig.1.  The quantitative effects of $ \kappa $  will be discussed in a future publication.
%  We expect that our  results  achieved in this paper could be tested in near future experiments in different systems.

%   {\bf Note added: } After finishing writing this preprint, we got to know that by adapting the idea proposed in \cite{dickej},
%   the $ Z_2 $ Dicke model \cite{chaos} was realized
%   in a previous and a recent experiments \cite{orbital} where the effective two "atomic" levels are the two momentum states of the cold atoms
%   in the optical lattice formed by
%   the cavity field and the off-resonant {\sl transverse } pumping Laser.
%   the effective energy scale is of the recoil energy $ E_r= \hbar^{2} k^{2}/2m \sim (1-10) kHz $ that
%   the $ Z_2 $ super-radiant phase was  observed experimentally.
%   We expect the $ U(1) $ super-radiant phase can also be realized in a similar experimental set-up.

   We thank  G. Cheng, T. Esslinger, B. Halperin, Han Pu, S. Sachdev, J. K. Thompson, V. Vultic, Jun Ye, X.L. Yu  and
   P. Zoller for very helpful discussions.
%J.Ye thank Han Pu for his
%hospitality during his visits at Rice university.
 JYe's research was
supported by NSF-DMR-0966413, NSFC-11074173, at KITP was supported
in part by the NSF under grant No. PHY-0551164.
% at KITP-C is
%supported by the Project of Knowledge Innovation Program (PKIP) of
%Chinese Academy of Sciences.
CLZ's work has been supported by National Keystone Basic Research
Program (973 Program) under Grant No. 2007CB310408, No. 2006CB302901
and by the Funding Project for Academic Human Resources Development
in Institutions of Higher Learning Under the Jurisdiction of Beijing
Municipality.

%   JYe's research at KITP was supported in part by the NSF under grant No. PHY-0551164, at KITP-C is
%   supported by the Project of Knowledge Innovation Program (PKIP) of Chinese Academy of Sciences.

\end{document}